\documentclass[galaxies,article,galaxies,accept,oneauthor,pdftex,10pt,a4paper]{mdpi}
\firstpage{1}
\articlenumber{x}
\doinum{10.3390/------}
\pubvolume{xx}
\pubyear{2016}
\copyrightyear{2016}
\externaleditor{Academic Editors: Jose L. G\'omez, Alan P. Marscher and Svetlana G. Jorstad}
\history{Received: 15 July 2016; Accepted: 23 August 2016; Published: date}

\usepackage{soul}
\usepackage{upgreek}
\usepackage{enumitem}
\usepackage{subfigure}
\usepackage{booktabs}
\usepackage{multirow}
\usepackage{longtable}
\usepackage{microtype}
\newcommand\myurl[1]{\changeurlcolor{black}\url{#1}\changeurlcolor{blue}}

\Title{Powers and Magnetization of Blazar Jets}
\Author{Marek Sikora}

\address[1]{Nicolaus Copernicus Astronomical Center, Polish Academy of Sciences/Bartycka 18, 00-716 Warsaw, Poland; sikora@camk.edu.pl}

\corres{Correspondence: }

\abstract{In this work I review the observational constraints imposed on the energetics and magnetisation of quasar jets, in the context of theoretical expectations. The discussion is focused on issues regarding the jet production efficiency, matter content, and particle acceleration. I show that if the ratio of electron-positron-pairs to protons is of order $15$, as is required to achieve agreement between jet powers computed using  blazar spectral fits and those computed using radio-lobe calorimetry, the magnetization of blazar jets in flat-spectrum-radio-quasars (FSRQ) must be significant. This result favors the reconnection mechanism for particle acceleration and explains the large Compton-dominance of blazar spectra that is often observed, without the need to postulate very low jet magnetization.}

\keyword{quasars; blazars; jets}

\begin{document}



\section{Introduction}

One of the biggest unresolved puzzles in the theory of active galactic nuclei (AGN) regards their ability to launch very
powerful, relativistic jets. According to several studies, the most powerful jets reach or even
exceed the associated accretion disk luminosities \cite{Rawlings,Fernandes,Punsly,SSKM,Ghisellini14}. Obviously, production of such jets cannot be treated as a
marginal byproduct of the accretion flow and most likely is governed by the Blandford-Znajek
mechanism \cite{BZ}, with the black holes possessing very large spins and
magnetic fluxes. The magnetic flux required to explain the production of the most powerful jets has
recently been found to agree with the maximum magnetic flux that can be confined on black holes by
the ram pressure of `magnetically-arrested-disks' (MAD) \cite{Narayan}. In recent years, the MAD scenario has been thoroughly investigated, and is now considered to be the likely remedy for the
production of very powerful \mbox{jets \cite{Tchek11,McKinney,Sikora-Begelman}}.
However, as numerical simulations
suggest, powers of jets launched in the MAD scenario depend not only on the spin and magnetic
flux, but also on the disk's geometrical \mbox{thickness \cite{Avara}}, with the jet power scaling approximately quadratically with all these quantities.

\begin{figure}[H]
\centering
\includegraphics[width=0.90\textwidth]{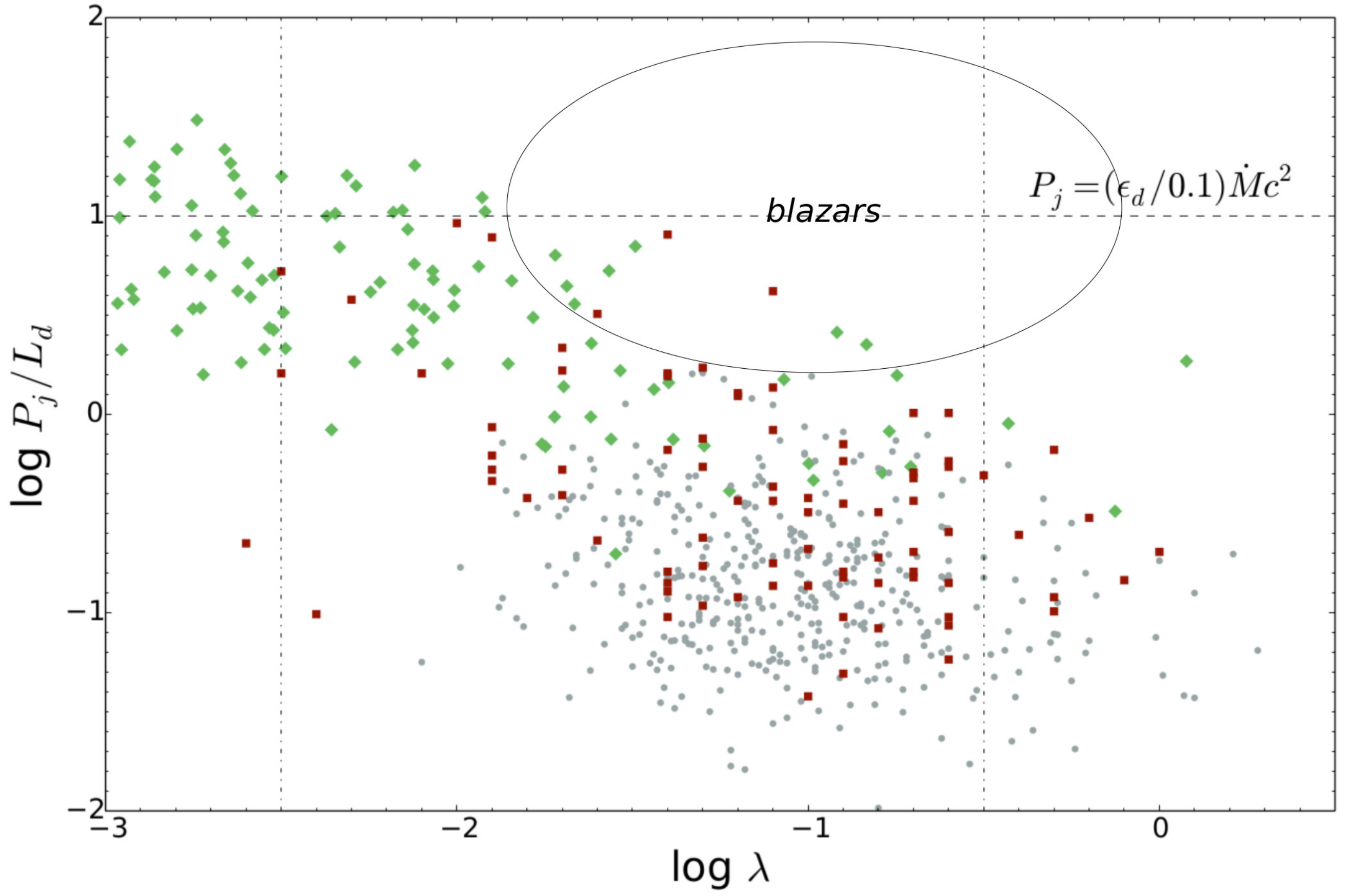}
\caption{\textls[-10]{$P_j/L_d$ ratio as a function of Eddington ratio $\lambda_{Edd} \equiv L_d/L_{Edd}$. Green diamonds---FRII Narrow-Line-Radio-Galaxies \cite{SSKM}; red squares---Broad-Line-Radio-Galaxies plus radio-loud quasars~\cite{SSL}; grey dots---FRII quasars~\cite{vanVelzen}; and the ellipse  marks the
approximate location of $\gamma$-ray selected FSRQs  assumed to have jets with  zero pair content \cite{Ghisellini14}. The vertical dashed lines mark the range of the Eddington ratio within  which the standard accretion disks are predicted to be radiatively efficient and geometrically thin. (The figure is adopted from Rusinek et al. 2016, in preparation \cite{Rusinek}).}}
\label{f1}
\end{figure}

According to standard accretion disk models, accretion disks become geometrically thick only
when they are advection dominated, i.e., for the Eddington-ratio
$\lambda_{Edd} \equiv L_d/L_{Edd}$ larger than \mbox{$\sim$0.3 \cite{Jaroszynski,Belob,Abramowicz}} or  smaller than $\sim$0.003, where $L_d$ is the disk luminosity and $L_{Edd}$ is the Eddington luminosity \cite{Ichimaru,Rees,Narayan-Yi,Stern-Laor}. Hence, one should not expect to observe
powerful jets in AGN with \mbox{$0.003 < \lambda_{Edd} < 0.3$.}
This theoretical expectation contradicts with the observational fact that there is no deficit of radio-loud
AGN in this range of $\lambda_{Edd}$. On the contrary, some studies show a trend of increasing jet production efficiency with
decreasing Eddington-ratio \cite{SSL,Rusinek} (see Figure~\ref{f1}).
A possible reason for the aforementioned contradiction is modulation of the accretion disk luminosity and jet production. Noting that the power of a jet calculated using the energy content of the
radio lobes is actually the time-averaged jet power, averaged over the source lifetime, and that according to the MAD scenario
modulation of a jet power at the base is driven by modulation of the accretion rate, modulation of
the jet production on time scales shorter than the lifetime of radio lobes will lead to modulation of
the ``apparent'' jet production efficiency and modulation of the Eddington ratio. For~example,
modulation of the accretion power by a factor 10 will cause the object to
have 10~times lower Eddington ratio and 10~times larger apparent jet production efficiency during its accretion rate minimum relative to that at its maximum, and for
a duty cycle ~1/2 its apparent jet production efficiency will be about 5~times larger than the real one. A natural driver of variability in the accretion rate and jet production
is viscous instabilities in accretion disks \cite{Janiuk1,Janiuk2}. Observational support for this hypothesis seems to come from the spatial modulation of the radio brightness distributions
seen in some large scale jets~\cite{Godfrey}.

However, the situation is complicated if we consider the jet power in $\gamma$-ray detected
FSRQs (flat-spectrum-radio-quasars) calculated by fitting their  broad-band  spectra assuming
the ERC (external-radiation-Compton) model for $\gamma$-ray production \cite{SBR}.
 The median of $P_j/L_d$ calculated by Ghisellini et al. (2014) for jets  in  FSRQs is $\sim$10,
which is $50$ times larger than the median of $P_j/L_d$ calculated   by van Velzen \& Falcke (2013)
for radio-selected FRII quasars using the radio-lobe calorimetry ({Willott et al. 1999} \cite{Willott}) (see Figure~\ref{f1}).
Studies of jet powers in blazars using calorimetry of their extended radio sources \cite{Kharb,Meyer}
allowed to verify whether such a difference is associated with selection of two different populations of quasars.
As it has been found, the $P_j/L_d$  median calculated using radio-lobe calorimetry
is still much smaller than the $P_j/L_d$ median calculated using the blazar spectral fits, but now they differ by a factor $\sim$16 \cite{Pjanka}. Whilst the larger averaged values of $P_j/L_d$ for jets in FSRQs than in FR II quasars, in both cases calculated using the radio-lobe calorimetry, can be explained
by selection procedures, the difference between medians of $P_j/L_d$ measured by the blazar spectral fits  and using radio-lobe calorimetry for these same samples must have a different explanation.
Following~possibilities have been considered:

\begin{enumerate}[leftmargin=*,labelsep=5mm]
\item[--] jet energy losses during propagation between the blazar zone and radio lobes (e.g., by the work
done against the external medium by reconfinement shocks which may change their sizes following
a jet power modulation by the central engine);

\item[--] overestimation of
jet power using the blazar models (e.g., by assuming zero pair content and/or a one zone model);

\item[--] a significant fraction of blazars may be hosted by young or short lived compact double radio sources. In  this case, the methods
of spectral decomposition and core subtraction adopted to use radio-lobe calorimetry in blazars
may  lead to underestimation of the lobe radio luminosities, and therefore underestimation of the jet power.
 \end{enumerate}

In this presentation I will assume that the main  reason for the discrepancy between jet power estimates in blazars
is overestimation of the jet power by the blazar models by assuming zero pair content.
It will be shown that $\sim$$15$ pairs per proton is enough to reconcile the difference
between the jet power calculated by the two methods and that with such a pair content the
magnetic reconnection mechanism is strongly favored as the energy source of blazar activity.


\section{Jet Powers}
I will assume that significant jet energy losses (relative to the initial jet power)  are taking place only via blazar radiation. In this case
the power of a jet leaving the blazar zone, $P_j$, is equal to the rate at which energy is delivered to
the radio lobes, $P_j^{(rl)}$, while the  power of the jet entering the blazar zone is
\begin{equation}
P_{j,0} = P_j^{(rl)} + P_{rad}
\end{equation}

Then the jet production efficiency is
\begin{equation}
\eta_j \equiv \frac {P_{j,0}}{\dot M_d c^2} = \frac {P_j^{(rl)} + P_{rad}}{\dot M_d c^2}
\end{equation}
where $P_{rad}$ is the rate of the jet energy losses dominated by the blazar radiation, $\dot M_d$ is the accretion rate, and all powers denoted by $P$ include contribution
from both a jet and a counter-jet (assumed to be equal). Now, noting that $P_j = P_B + P_p + P_e = P_j^{(rl)}$,
and that for FSRQs typically  $P_e \ll P_B$ \cite{Ghisellini14}, we have
\begin{equation}
P_p = P_j^{(rl)} - P_B
\end{equation}
where $P_B$, $P_p$, and $P_e$ are the magnetic, proton, and electron energy fluxes, respectively.
Since for leptonic radiative models $P_p = (n_p/n_e) P_{p,n_p=n_e}$, we can find
\begin{equation}
\frac {n_e}{n_p} = \frac {P_{p,n_p=n_e}}{P_j^{(rl)} - P_B}
\end{equation}
where $P_{p,n_p=n_e}$ is the proton energy flux assuming proton-electron plasma
(zero pair content), \mbox{$n_e \equiv n_{e^+} + n_{e^-}$}, and the number of pairs per proton is
$n_{pairs}/n_p \equiv n_{e^+}/n_p = (n_e/n_p -1)/2$.

\textls[-10]{For $P_{p,n_p=n_e} \simeq 2.5 \times 10^{46} \, {\rm ergs\cdot s^{-1}}$,
$P_B \simeq \kappa_B 10^{45} \, {\rm ergs\cdot s^{-1}}$, $P_{rad} \simeq 2.0 \times 10^{45} \, {\rm ergs\cdot s^{-1}}$, and \mbox{$\dot M_d c^2 = L_d/\epsilon_d = 3.16 \times 10^{46}/(\epsilon_d/0.1)  \, {\rm ergs\cdot s^{-1}}$} which
are averages obtained by  Ghisellini et al. (2014)~\cite{Ghisellini14} for a sample of $191$ $\gamma$-ray detected FSRQs,
and for \mbox{$P_j^{(rl)} \sim 0.1 \times P_{p,n_p=n_e} \sim 2.5 \times 10^{45} \, {\rm ergs\cdot s^{-1}}$}
found by Pjanka et al. \cite{Pjanka}, Equations (1)--(4) give}

\bigskip
\centerline {$P_p \sim 0.83 \times 10^{45} \,  {\rm ergs\cdot s^{-1}}$\,  ($\to \sigma \simeq P_B/P_p \sim 2$)}

\smallskip
\centerline {$P_{j,0} \sim 4.5 \times 10^{45} \,{\rm ergs\cdot s^{-1}}$}

\smallskip
\centerline {$\eta_j  \sim 0.13 (\epsilon_d/0.1) $}

\smallskip
\centerline { $n_e/n_p \sim 30$ ($\to n_{pairs}/n_p \sim 15$)}

\bigskip
\noindent
where $\epsilon_d$ is the accretion disk radiation efficiency, $\kappa_B \equiv (p_B+u_B')/u_B'$,
$u_B'$ is the magnetic energy density, and $p_B$ is the magnetic pressure (not included in Ghisellini et al. \cite{Ghisellini14}).
For purely toroidal magnetic field $\kappa_B = 2$, while for turbulent, isotropic magnetic field $\kappa_B = 4/3$. We assumed in our estimations $\kappa_B=5/3$, which corresponds to equal energy densities of toroidal and turbulent, isotropic magnetic field components.
\section{Jet Magnetization}

Jets which are launched by the Blandford-Znajek mechanism are initially strongly
Poynting-flux dominated. Following the conversion of the magnetic energy
flux to the kinetic energy flux the jets are accelerated \cite{Sikora05,Tchek09,Lyubarsky,Komissarov} reaching bulk Lorentz factors on sub-parsec/parsec scales ranging from a few up to tens.
Whether their magnetization parameter $\sigma$ defined to be the magnetic-to-kinetic
energy flux ratio drops to unity or much smaller values is still debated.
Whereas blazar models allow a determination of the magnetic energy flux, the kinetic
energy flux depends on the unknown proton content (fits of blazar spectra by leptonic models  provide us
information about a number of electrons plus positrons, not about the number of protons). As was discussed in the preceding
sections, pure electron-proton models predict jet powers which, for a large fraction  of gamma-ray detected blazars, are not  reachable even by the MAD models
with maximal jet production efficiency, and have problems accounting for jet powers calculated using the lobe
energetics which are more than 10 times smaller than the jet powers calculated
using the blazar spectral fits. On the other hand, reducing the jet power by postulating
$n_p \ll n_e$ implies much larger efficiency of radiative processes in blazars than in case of e-p models. Specifically, such efficiency is
\begin{equation}
\epsilon_j \equiv \frac {P_{rad}}{P_B + P_p + P_{rad}}
\end{equation}
which for $P_p=P_p(n_p=n_e)$ and averaged values of powers taken from Ghisellini
et al. \cite{Ghisellini14} gives $\epsilon_j \simeq 0.07$, while for
$P_p \simeq P_j^{(rl)} - P_B$ gives $\epsilon_j \simeq 0.45$. Such high radiative efficiency and magnetization $\sigma \simeq \kappa_B P_B/P_p \simeq 2$ favors the reconnection of magnetic fields  as the mechanism powering the  blazar radiation \cite{Sironi}.
Assuming that a change of the jet speed within the blazar zone is negligible, one can find that
at the entrance to the blazar zone the jet magnetization is $\sigma_0 \sim P_{B,0}/P_K \simeq (P_B + P_{rad})/P_p \sim 4.4$.

The ERC model of $\gamma$-ray production predicts the ERC-to-synchrotron luminosity peak ratio
\begin{equation}
q= \frac{L_{ERC,peak}}{L_{syn, peak}} \simeq \frac {u_{ext}'}{u_B'} \simeq
\frac {1+\sigma}{2\sigma} \,  \frac {\zeta \epsilon_d} {\eta_j} \,
\Gamma^2 \, (\theta_j \Gamma)^2
\end{equation}
where $\zeta$ is the fraction of disk radiation reprocessed into broad emission lines, $\theta_j$
is the half-opening angle of a jet, and $u_{ext}'$ and $u_B'$ are energy densities of external
radiation field and of magnetic field, respectively \cite{Janiak}.
For typical blazar parameters and  $n_p=n_e$ this gives the observed values of the Compton-dominance $q \gg 1$ only for  $\sigma \ll 1$ \cite{Nalewajko-SB,Janiak}.
For $n_p \ll n_e$ the jet production
efficiency $\eta_j$ is much lower and therefore $q \gg 1$ can be reached even for $\sigma > 1$.

\section{Particle Acceleration and Spectral Peaks}

The shape of the electron/positron energy distribution required to reproduce the observed spectra of high
Compton-dominance blazars naturally results from injection of
electrons with the break at energy  corresponding roughly with
the average electron  injection energy \cite{Janiak}.
Assuming that each electron is accelerated within
the blazar zone only once, that energy is given by
\begin{equation}
\bar \gamma \simeq \frac{\epsilon_e \epsilon_{diss} P_{j,0}}{m_e c^2 \dot N_e \Gamma}
\end{equation}
where $\dot N_e = (n_e/n_p) \dot N_p = (n_e/n_p) \dot M_p/m_p$, $\dot M_p$ is
the proton mass flux in a jet, and $\epsilon_e$ is the fraction of dissipated energy channeled to relativistic electrons and positrons.
For the jet power dominated by magnetic and proton energy fluxes,
\begin{equation}
P_{j,0}=(1+\sigma_0)(\Gamma_0-1) \dot M_p c^2;  \,\,\,
P_j=(1+\sigma)(\Gamma-1) \dot M_p c^2
\end{equation}
and then in case of $\Gamma=\Gamma_0$,
\begin{equation}
\epsilon_{diss} \equiv \frac {P_{j,0} - P_j} {P_j} =  \frac {\sigma_0 - \sigma}{1+\sigma_0}
\end{equation}

For $\Gamma_0 \gg 1$ these  equations give
\begin{equation}
\bar \gamma = (\sigma_0 - \sigma)\, \epsilon_e \, \frac {n_p m_p} {n_e m_e}
\end{equation}

Electrons with such energy are expected to produce synchrotron and ERC
luminosity peaks. In~order to reproduce their location in the electromagnetic spectrum, with the synchrotron emission peaking at around $10^{13}$ Hz and the ERC emission peaking at around
$10^{22}$ Hz {\cite{Giommi}, the electron energy should be of the order of 100 \cite{GT-FSRQ}. For the parameters derived above ($\sigma_0 \simeq 4.4$, $\sigma \simeq 2$, and $n_e/n_p\simeq 30$), and noting that
the fraction of dissipated energy channeled to electrons/positrons is expected to be
$0.5 < \epsilon_e < 1$, this condition is satisfied.

\section{Discussion and Conclusions}

The jet powers in FR II quasars calculated using radio-lobe calorimetry \cite{vanVelzen} are a factor \mbox{$\sim$50 smaller} than jet powers calculated using spectral fits of $\gamma$-ray detected FSRQs when assuming $n_p=n_e$ jet plasma \cite{Ghisellini14} (see Figure~\ref{f1}).
Measurements of radio luminosities of extended radio-structures in some  FSRQs \cite{Kharb,Meyer} has enabled a comparison of the results of the two jet power measurement techniques by applying both techniques to the same objects.
As Pjanka et al. \cite{Pjanka} showed, the average difference for the cross-matched samples is  smaller, but still significant, by a factor $\sim$$16$. Noting that the power calculated by Ghisellini et al.  includes
the blazar radiation, whilst the power calculated using the radio lobe energetics does not,
the difference is reduced to a factor $\sim$$8$.  Assuming that this remaining difference is due to
overestimation the blazar power by Ghisellini et al. by ignoring pair plasma in a jet, we found that
on average:

\begin{itemize}[leftmargin=*,labelsep=5mm]
\item the pair content: $n_{pairs}/n_p \sim 15$;
\item the jet production efficiency:  $\eta_j \simeq 0.14 \, (\epsilon_d/0.1)$;
\item the magnetization at entrance and exit of blazar zone is $\sigma_0 \simeq 4.4$ and $\sigma \simeq 2$, respectively;
\item the average energies of accelerated electrons/positrons correspond to typical observed locations
of synchrotron and $\gamma$-ray luminosity peaks provided the acceleration is powered by magnetic  reconnection operating in a jet at a distance corresponding to spatial extension of the BLR.
\end{itemize}

The pair content obtained above is larger by a factor $3/2$ than the maximal one, \scalebox{.95}[1.0]{$(n_{pairs}/n_p)_{max} \sim 10$},
above which the jet is expected to be  efficiently decelerated due to
the Compton-rocket effect \cite{GT-Rocket}. However
this limit was obtained by assuming an isotropic distribution of the external diffuse radiation field
and neglecting Klein-Nishina effects and therefore can be relaxed somewhat if some level of flattening of the BLR is considered, and taking into account Klein-Nishina recoil effects \cite{Moderski}.

\noindent
Noting that the MAD scenario predicts production of jets with powers
\begin{equation}
P_{j,0} \simeq a^2 \, (H/R)^2 \, \dot M_d c^2
\end{equation}
and assuming that spins $a$ in $\gamma$-ray selected quasars are close to their maximum value, $\sim$1,
the value of $\eta_j \sim 0.14$ obtained above may simply result from suppression of jet production efficiency due to having geometrically thin disks in quasars, i.e., with $H/R \ll 1$.
This  requires $H/R \sim 1/3$, which is much larger  than predicted by standard accretion disk models
\cite{Shakura,Novikov-Thorne}. However, as some  studies indicate, real disks  can be thicker than the standard disks \cite{BegPri,Rozanska,BAR}.
Then the $\sim$$7$ times lower jet production efficiency  in radio selected FRII quasars than in
$\gamma$-ray detected FSRQs may result from on average \mbox{$\sim$$\sqrt{7}$ times} lower BH spins in the entire FRII quasar population than in the $\gamma$-ray selected one. And~this can be explained by noting, that in
$\gamma$-ray loud quasars jets are more powerful and therefore are expected to be more relativistic.
This interpretation is supported by the fact that only about $9$\% of the FSRQs are found to be $\gamma$-ray loud \cite{Linford}.

\vspace{6pt}
\acknowledgments{I thank Leith Godfrey, Krzysztof Nalewajko, Katarzyna Rusinek and  Andrzej Zdziarski for useful comments. This work
was partially supported by the Polish National Science Centre grant 2013/09/B/ST9/00026 and by the conference organizers.}

\conflictofinterests{{The} author declares no conflict of interest.}

\bibliographystyle{mdpi}

\renewcommand\bibname{References}

\end{document}